\documentclass[aps,prd,twocolumn]{revtex4}  
\pdfoutput=1
\usepackage{graphicx}
\usepackage{textcomp}
\usepackage{bm}
\usepackage{times}
\usepackage{color}
\usepackage{graphics}
\usepackage{hyperref}
\usepackage{setspace}
\usepackage{slashed}
\usepackage{paralist}
\usepackage{bbm}
\usepackage{ulem}
\usepackage{enumitem}
\usepackage{color}
\usepackage{slashed}
\usepackage{comment}

\usepackage{rotating}
\usepackage{framed}

\usepackage{placeins}

\usepackage{cancel}

\usepackage{amsmath} % AMS Math Package
\usepackage{amsthm} % Theorem Formatting
\usepackage{amssymb}	% Math symbols such as \mathbb
\let\baraccent=\= % rename builtin command \= to \baraccent
\renewcommand{\=}[1]{\stackrel{#1}{=}} % for putting numbers above =

\theoremstyle{definition}

\theoremstyle{remark}

\usepackage{bbm}

\hypersetup{
    pdfnewwindow=true,      % links in new window
    colorlinks=true,       % false: boxed links; true: colored links
    linkcolor=black,          % color of internal links
    citecolor=blue,        % color of links to bibliography
    filecolor=blue,      % color of file links
    urlcolor=blue           % color of external links
}

\makeatletter
\newcommand\xleftrightarrow[2][]{%
  \ext@arrow 9999{\longleftrightarrowfill@}{#1}{#2}}
\newcommand\longleftrightarrowfill@{%
  \arrowfill@\leftarrow\relbar\rightarrow}
\makeatother

\begin{document}
%%%%%%%%%%%%%%%%%%%%%%%%%%%%%%%%%%%%%%
\title{\Large {{\bf{Gamma Lines from the Hidden Sector}}}}
\author{Pavel Fileviez P\'erez$^{1}$, Clara Murgui$^{2}$}
\affiliation{$^{1}$Physics Department and Center for Education and Research in Cosmology and Astrophysics (CERCA), 
Case Western Reserve University, Rockefeller Bldg. 2076 Adelbert Rd. Cleveland, OH 44106, USA \\
$^{2}$Departamento de F\'isica Te\'orica, IFIC, Universitat de Valencia-CSIC, E-46071, Valencia, Spain}
\begin{abstract}
We discuss the visibility of gamma lines from dark matter annihilation. We point out a class of theories for dark matter which predict the existence of gamma lines with striking features. In these theories, the final state radiation processes are highly suppressed and one could distinguish easily the gamma lines from the continuum spectrum. 
We discuss the main experimental bounds and show that one could test the predictions 
for gamma lines in the near future in the context of simple gauge theories for dark matter. 
\end{abstract}
\maketitle

%%%%%%%%%%%%%%%%
\section{Introduction}
%%%%%%%%%%%%%%%%
There are many evidences today of the existence of dark matter (DM) in the Universe, and thanks to many impressive experiments 
we know that the relic abundance must be, approximately, $\Omega_{\rm{DM}} h^2 \approx 0.12$~\cite{CMB}. 
Unfortunately, we only know the amount of dark matter we need to be consistent with the experiments, i.e. we know nothing about its nature.
Thanks to the imagination and creativity of our physics community, we have a large list of possible candidates for the cold dark matter.

 \begin{figure}[b]
\centering
\includegraphics[width=0.8\linewidth]{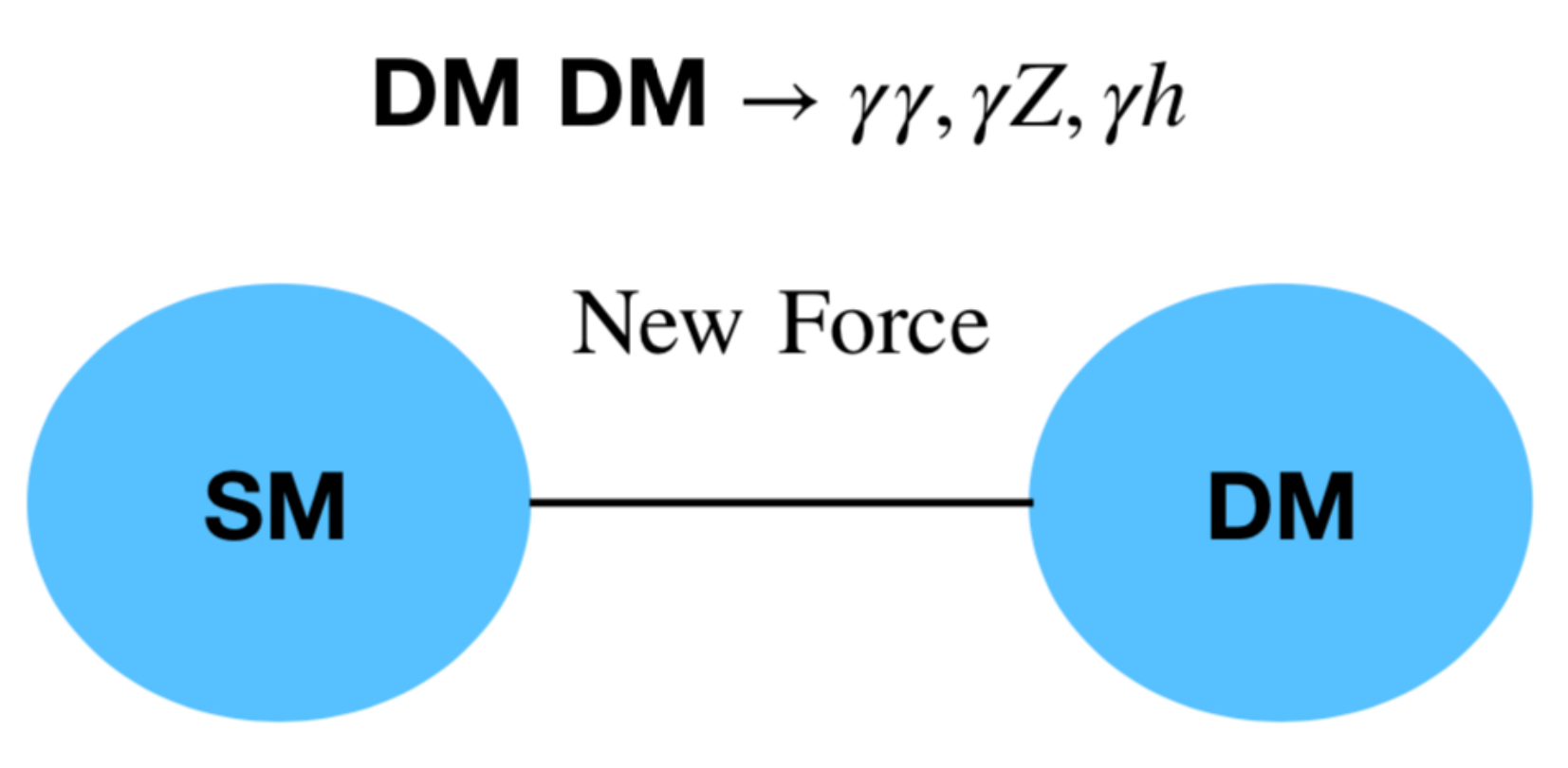} 
\caption{Gamma Lines from the Hidden Sector.}
\label{Our-goal}
\end{figure}

From the theoretical point of view, the weakly interacting massive particles (WIMPs) are still, in our opinion, one of the best candidates for cold dark matter.
The existence of WIMPs is predicted in many extensions of the Standard Model (SM) and one can understand most of their properties.
In theories for physics beyond the SM, one can explain the stability of WIMPs, compute their relic density and understand the predictions for direct and indirect dark matter experiments.

We have several ways to look for signatures from particle dark matter candidates: a) Direct detection, b) Indirect detection, and c) Signals at colliders. In direct detection experiments, one looks mainly for signatures due to the scattering of WIMPs with nucleons or electrons. However, these processes could be highly suppressed or not even possible. At colliders, one could observe missing energy signatures 
related to dark matter but, since one cannot probe the stability of the DM, one needs results from other experiments to confirm the existence of DM.

\begin{figure}[h]
\centering
\includegraphics[width=0.8\linewidth]{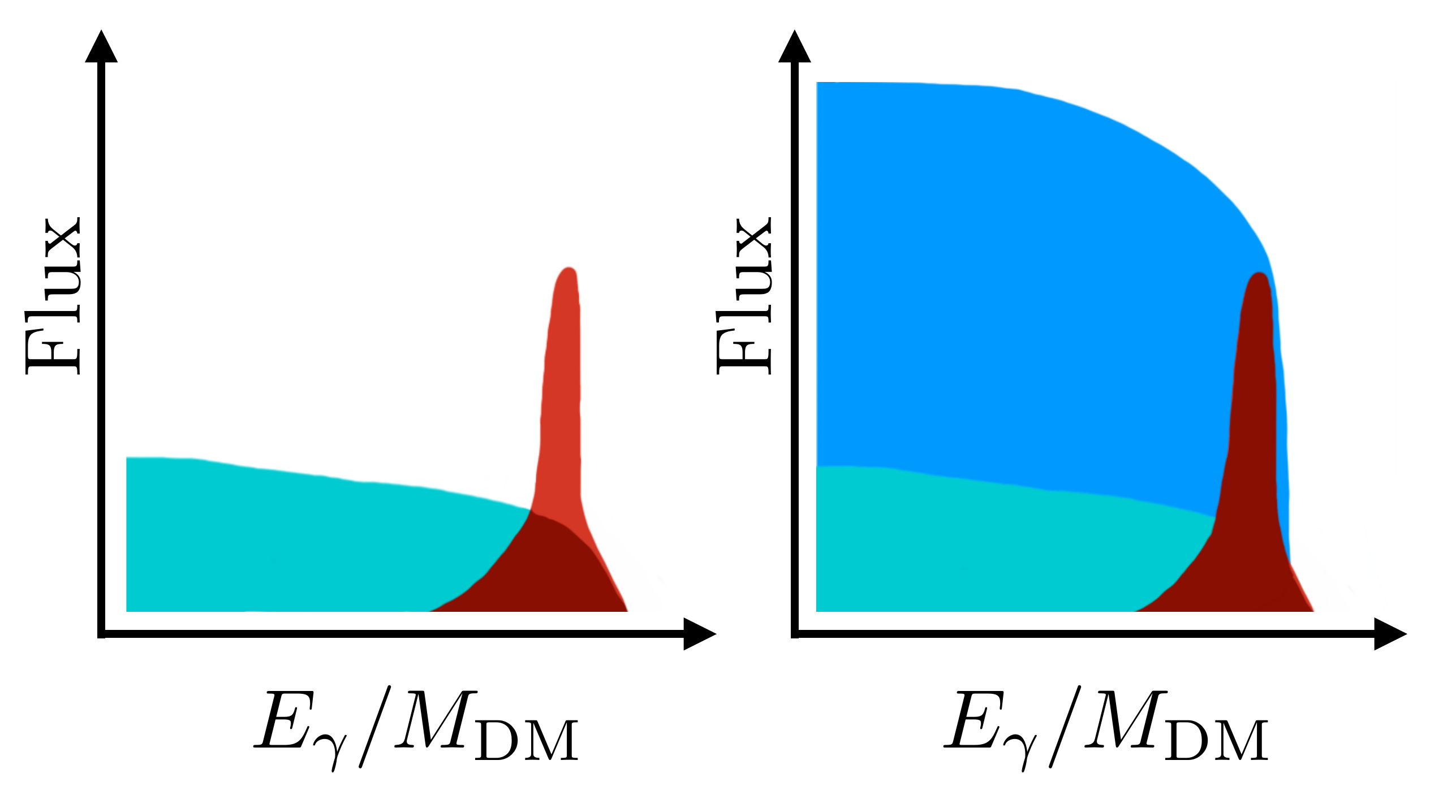} 
\caption{Naive illustration of the spectrum for gamma rays from dark matter annihilation in two scenarios. 
In the left-panel we show the scenario where the gamma line is above 
the continuum when processes such as the final state radiation processes are highly suppressed, 
while in the right-panel we show the scenario where the gamma line is not visible.}
\label{line-continuum}
\end{figure}

In indirect detection experiments, one looks for signals from DM annihilation into SM particles. For example, one looks for the DM annihilation 
into leptons, quarks, neutrino lines or gamma lines. In general, we do not know if the DM couples to any SM particle and then we cannot be 
sure about this possibility. The gamma lines from DM annihilation are, perhaps, the most striking signatures which are free of any astrophysical backgrounds 
or issue about modelling the propagation of the annihilation products. Gamma lines could be generated even if the DM does not couple to any SM particle; 
one only needs some interactions with electric charged particles. If one day these gamma lines are observed in current or future gamma-ray telescopes, one could make a map of the DM halo in a given galaxy. Unfortunately, these processes are quantum mechanical and often suppressed.
See Ref.~\cite{Bringmann:2012ez} for a review about the predictions of gamma lines in different models.
 
In this article, we investigate the predictions for gamma lines from dark matter annihilation in models where the dark matter lives in a Hidden sector.
See Fig.~\ref{Our-goal} for a simple illustration of our main goal. The fact that we do not see any signal in current dark matter experiments may be suggesting that the DM 
does not have any direct coupling to the SM fields and lives in a Hidden sector. Now, the observation of gamma rays depends on two main factors: a) the annihilation 
cross section should be large to be able to see a signal at gamma-ray telescopes, b) the continuum spectrum generated by other physical processes must 
be suppressed in order for the gamma line to be visible. In a given dark matter model, one can have physical processes such as final state radiation emission 
which may spoil the visibility of gamma lines. 

In Fig.~\ref{line-continuum}, we show a naive illustration of the spectrum for gamma rays from 
dark matter annihilation in two scenarios. In the left-panel we show the scenario where the gamma line is above the continuum when processes 
such as the final state radiation processes are highly suppressed, while in the right-panel we show the scenario where the gamma line is not visible.
Of course, as we will show later, one needs a very good energy resolution to be able to resolve the gamma lines from dark matter annihilation.   

In this Letter, we point out a class of theories for dark matter which predict the existence of gamma lines with striking features. 
In these theories, the final state radiation processes are highly suppressed and one could distinguish easily the gamma line 
from the continuum spectrum. We discuss the main experimental bounds and show that one could test the predictions 
for gamma lines in the near future in the context of simple gauge theories for dark matter. 

This article is organized as follows: In section II we discuss the main features of gamma lines in different dark matter models.
In section III we discuss the predictions of gamma lines in two different models and discuss the visibility of these gamma lines, 
while in section IV we show some numerical examples to illustrate the visibility of gamma lines. Our main findings are summarize in section V.

%%%%%%%%%%%%%%%%%%%%%%%%%%%%
\section{DM ANNIHILATION INTO GAMMA RAYS}
%%%%%%%%%%%%%%%%%%%%%%%%%%%
In Fig.~1, we show a simple illustration of the possible connection between the Hidden and SM sectors. The Hidden sector contains, by definition, only new fields which are not charged under the Standard Model gauge symmetry. One can imagine many ways to define the connection between the DM and SM sectors. This connection 
defines the predictions for DM direct and indirect detection experiments. Unfortunately, the current DM direct detection 
bounds tell us that, soon, one may reach the so-called neutrino floor, making more difficult the detection 
of DM in direct detection experiments. However, one could, in general, expect 
the existence of gamma lines due to the DM annihilation into gamma rays, i.e. $\rm{DM} \ \rm{DM} \to \gamma \gamma, \gamma Z, \gamma h$.
These signatures are very striking and almost background free. 

The energy of the gamma lines is given by
\begin{equation}
E^\gamma_i = M_{\rm{DM}} \left(1 - \frac{M_i^2}{4 M_{\rm{DM}}^2} \right),
\label{Eq1}
\end{equation}
where $M_i=0,M_Z,M_h$ for the $\rm{DM} \ \rm{DM} \to \gamma \gamma, \gamma Z, \gamma h$ annihilation channels, respectively.
Therefore, one would know the dark matter mass if one of these gamma lines could be observed in the near future. 

Now, in a given dark matter model, one can have final state radiation processes such as $\rm{DM} \ \rm{DM} \to \rm{SM} \ \rm{SM } \ \gamma$ which may spoil the visibility of the gamma lines.
The maximal energy of the photon in the final state radiation processes is given by
\begin{equation}
E^\gamma_{\rm{max}} = M_{\rm{DM}} \left(1 - \frac{M_{\rm{SM}}^2}{ M_{\rm{DM}}^2} \right),
\label{Eq2}
\end{equation}
where $\rm{M}_{\rm{SM}}$ is the mass of a Standard Model electric charged field. We note that, typically, this process occurs at tree level while the gamma lines discussed above are quantum mechanical processes.
  
Therefore, the relevance of the final state radiation with respect to the gamma lines will be crucial to determine whether they can be observed or not. The fact that they can be distinguished from the continuum spectrum will depend on the features of the model, as well as the nature of the DM candidate. Let us start discussing different DM candidates:

\begin{itemize}

\item Scalar DM

In the scenario where the DM is spinless, $\phi$, one always has the Higgs portal term $\phi^\dagger \phi H^\dagger H$.
Therefore, the annihilation into two SM fermions is, in general, not suppressed. The final state radiation processes where 
one emits a photon from one of the SM fermions coming from the DM annihilation are also not suppressed. 
Since the final state radiation processes are tree level processes and the gamma lines are quantum mechanical 
ones, the visibility of the gamma line is in general spoiled. If the mediator between the DM and SM sector is a gauge 
force, the DM annihilation into two SM fermions is velocity suppressed, and the same occurs with the gamma lines.   
Therefore, in the simplest DM scenario with scalar DM, one could not distinguish the gamma lines from the continuum spectrum.
For the study of gamma lines in scalar DM models see Refs.~\cite{Gustafsson:2007pc,Garcia-Cely:2015khw,Duerr:2015bea,Duerr:2015aka}.

\item Dirac DM 

One can imagine many possible scenarios for Dirac DM. Here, we will focus our discussions in the context 
of simple gauge theories for DM. In this context, the final state radiation processes, $\bar{\Psi} \Psi \to \bar{f} f \gamma$, 
are not suppressed and, generally, one could not distinguish the gamma lines generated by DM annihilation from the continuum spectrum. One could consider, however, models where the DM annihilation takes place through a scalar mediator. In these scenarios, the Higgs portal term between the mediator and the SM Higgs will determine the annihilation channels and one will have a similar situation as in the scalar DM scenario mentioned above. For the predictions of gamma lines in models for Dirac DM see Refs.~\cite{Jackson:2013pjq,Jackson:2009kg}.

\item Majorana DM

One can have simple gauge theories for dark matter where one has gamma lines with striking  features when the DM is a Majorana fermion.
In these scenarios, the DM annihilation through the new gauge boson in the theory is velocity suppressed and then the final state 
radiation processes are highly suppressed. If the amplitudes relevant for the gamma lines are not velocity suppressed, then 
we can hope to observe the gamma lines, since they can be distinguished from the continuum spectrum. For the study of the gamma lines in models 
with a Majorana dark matter see Refs.~\cite{Bergstrom:1997fh,Bergstrom:2005ss,Cirelli:2015bda,Garcia-Cely:2015dda,Duerr:2015vna}.

\end{itemize}

See also Refs.~\cite{Bergstrom:2004nr,Bergstrom:2004cy} for the study of gamma lines in other dark matter models.

In this article, we will discuss the simplest gauge theories for Majorana DM where the final state radiation processes are highly suppressed and it is possible to observe the gamma lines from DM annihilation. These DM theories have a few features: 
a) We have a Hidden sector which contains the DM candidate, b) The connection between the DM and SM sectors is defined mainly by the new gauge force, c) The SM fermions are charged under the new force,  $U(1)_X$, which we assume that it is an Abelian force for simplicity.  

Therefore, in our framework we start from a general abelian extension of the gauge symmetry of the SM, 
$$ SU(3)_c \otimes SU(2)_L \otimes U(1)_Y \otimes U(1)_X,$$
where the $U(1)_X$ is the new abelian force which could be, for instance, local baryon ($B$) or lepton ($L$) number, or linear combinations of them such as $B-L$, etc.. We assume that this new symmetry is spontaneously broken at some scale. The details on how the symmetry is broken are irrelevant for the discussion on the study of gamma lines, but we point out that the theory could undergo symmetry breaking if, for instance, there is a scalar charged under the new symmetry with a non-zero VEV. 

The Lagrangian relevant for our discussions is given by
\begin{equation}
\begin{split}
{\cal L} \supset & g^\prime  \, \bar{\chi} \gamma^\mu \gamma_5 \chi Z^\prime_\mu + g^{\prime} \bar{f} (n_V^f \gamma^\mu + n_A^f  \gamma^\mu \gamma_5)f Z^\prime_\mu \\
&+ \bar{f}\gamma^\mu (g_{V}^f+g_{A}^f\gamma_5) f Z_\mu, 
\end{split}
\end{equation}
where $g'$ is the gauge coupling associated to the new force. We note that, once the $U(1)_X$ is broken, the DM field, $\chi$, becomes a Majorana fermion and couples axially to the new gauge boson $Z_\mu^{'}$. The couplings $n_i$ refer to the interactions between the new mediator and any fermion, $f$, of the theory; whereas the couplings $g_i$ parametrize how these fermions interact with the SM $Z$ boson. The subindex $V$ and $A$ refer to the vector and axial components of the new gauge interaction, respectively. We assume that the theory has the right fermion content in order to be anomaly free. However, henceforth, we will only consider the SM fermions without loss of generality, since they will give the most relevant contributions to the cross section.

\begin{figure}[h]
\centering
\includegraphics[width=0.45\linewidth]{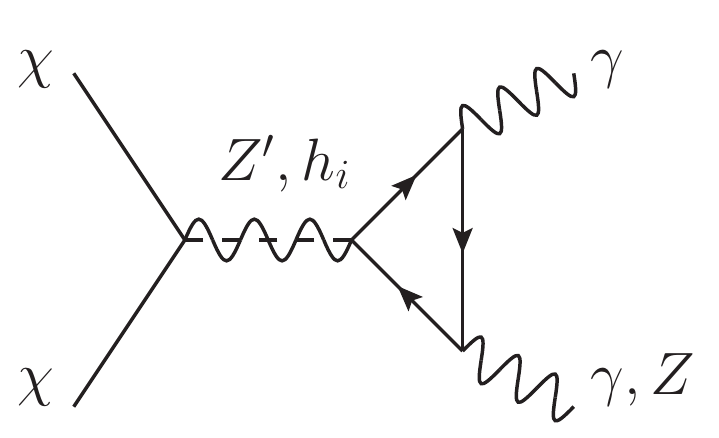} 
\includegraphics[width=0.45\linewidth]{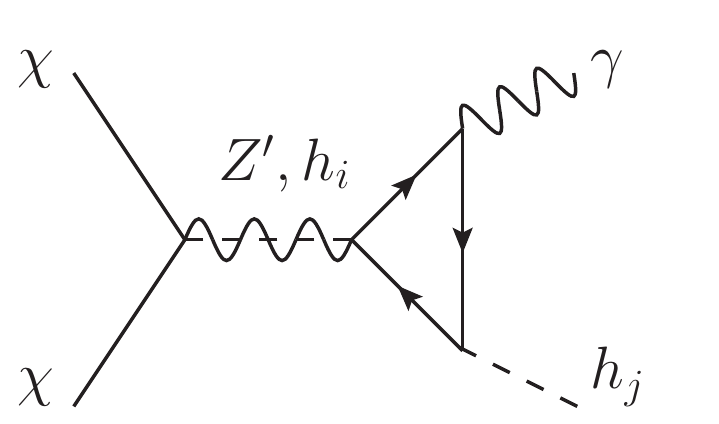} 
\caption{Dark matter annihilation into gamma rays in gauge theories.}
\label{graphs}
\end{figure}

The possible topologies through which the Majorana DM candidate can annihilate into gamma lines are shown in Fig.~\ref{graphs}. The fermions in the loop are SM fermions. We note that, in general, the Higgses present in the theory can mediate these processes, but these contributions are not relevant for our discussions because they are velocity suppressed and, therefore, neglectable. In the following, we will only consider the cases mediated by the new vector boson.

%%%%%%%%%%%%%%%%%%%%%%%%%%%%
%%%%%%%%%%%%%%%%%%%%%%%%%%%%

\begin{figure}[h]
\centering
\includegraphics[width=0.45\linewidth]{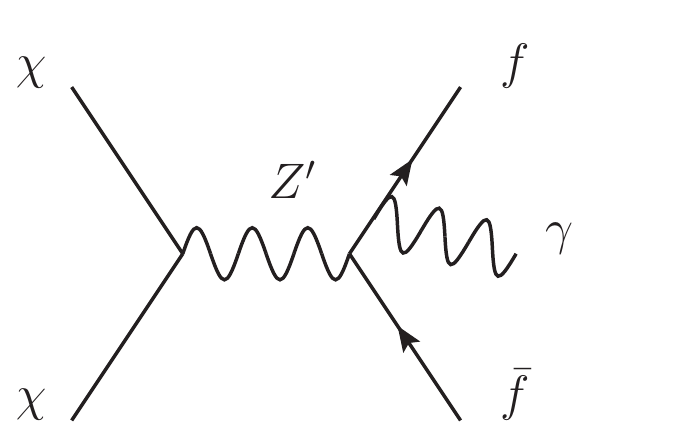} 
\includegraphics[width=0.45\linewidth]{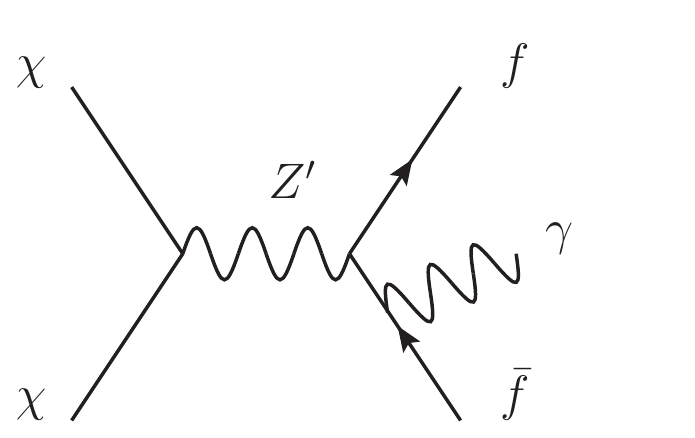} 
\caption{Contributions to the final state radiation processes.}
\label{graphsFSR}
\end{figure}

Again, we would like to emphasize that, in order to observe these gamma lines in indirect searches, the cross section for the gamma lines must be much larger than 
the cross section for the final radiation processes. See Fig.~\ref{graphsFSR} for the contributions to the final state radiation processes. In the case we study here, the squared amplitude for the FSR can be written in the following way
\begin{equation}
|{\cal M}|^2_\text{FSR} = \frac{M_f^2}{M^2_{Z'}}A +  v^2 B + {\cal O}(v^4).
\label{Eq3}
\end{equation}
Then, the FSR is suppressed either by the mass of the $Z'$ or the velocity of the dark matter candidate. 
See appendix B for the explicit form of the coefficients $A$ and $B$ in the above equation.
Therefore, one can have striking predictions for Majorana dark matter candidates in the context of these gauge theories. 
We note that, on the other hand,  the DM-nucleon elastic scattering is also velocity suppressed.
Therefore, direct searches will be, in general, suppressed and close or below the neutrino floor limit.

In Table~\ref{Tab:combinations}, we show the different combinations of couplings which can give us non-zero contributions to gamma lines. 
Let us discuss some main features of the gamma lines:
\begin{table}[h]
\begin{tabular}{ | c | c | c | c |}
\hline
Gamma Lines & $ \phantom{g_V g_V} \gamma \gamma \phantom{g_V g_V}$ & $ Z \gamma  $ & $\phantom{g_V} h \gamma  \phantom{g_V}$ \\
 \hline
 Couplings & $n_A^f$ & $\phantom{g_V} \xcancel{n_V^f \cdot g_{A}^f} ,  n_A^f \cdot g_{V}^f \phantom{g_V}$ & ${n_V^f}$  \\ 
 \hline
\end{tabular}
\caption{Combination of couplings which generates a non-zero effective vertex relevant for the different gamma lines. }
\label{Tab:combinations}
\end{table}

\begin{enumerate}[label=(\alph*)]
\item the term proportional to $n_V^f$ entering in the DM annihilation into $Z\gamma$ is insensitive to the mass of the fermions running inside the loop. The amplitude is proportional to 
$${\cal A}   \propto  \sum_f N_c^f Q_f n_V^f g_{A}^f  , $$
which is zero for any well defined theory, ensured by the anomaly cancellation. In the same way, any combination of the charges coming from the loop which is insensitive to the fermion masses will be zero, i.e. the following combinations:
$$ {\cal A}  \propto  \sum_f N_c^f Q_f n_A^f g_{V}^f  \,   \text{  and  } \, \propto \sum_f N_c^f Q_f^2  n_A^f . $$
\item The annihilation cross section into $h\gamma$ is velocity suppressed and we can neglect it. 
\end{enumerate}

We would like to emphasize that, for any vector mediator connecting the fermionic dark sector with the fermions of the theory, only the theories that accommodate fermions that couple axially to the new gauge boson will be relevant for detection of gamma lines, as summarized in Table~\ref{Tab:combinations}. This result has dramatic consequences for anomaly free theories such as $U(1)_{B-L}$,  $U(1)_{B-3L_i}$ or $U(1)_{L_i - L_j}$, in which the mediator only couples vectorially to the fermions of the theory and, therefore, no relevant gamma line for indirect searches is expected. 
In spite of this, in the next section, we will introduce a set of well-motivated theories which satisfy the above theoretical requirements of being testable in the context of indirect searches. 
 
 %%%%%%%%%%%%%%%%%%%%%%%%%%
\section{DM Theories and Visible Gamma Lines}
%%%%%%%%%%%%%%%%%%%%%%%%%%
%
In the previous section, we have mentioned that the connection between the SM and the Hidden sectors will be defined by a new gauge force.
In this section, we will discuss two main classes of gauge theories for dark matter where the final state radiation processes do not overcome the gamma line signals:
\begin{itemize}
\item {\bf{Model I}}:  One can have a simple anomaly free theory, adding three right handed neutrinos, when we gauge 
a linear combination of the weak hypercharge and the baryon minus lepton, $B-L$, quantum numbers, i.e. 
$$ U(1)_X :  X=aY+b(B-L).$$
We note that the new gauge boson will have axial couplings to the SM fermions. As we have previously discussed, this is the crucial ingredient for a theory with spin one mediator to be testable in indirect searches. 
The relevant interactions for the annihilation into gamma lines are given by the following Lagrangian:
\begin{eqnarray}
{\cal L} &\supset & \frac{g'}{3}\bar{d}\slashed{Z'} \left(-\frac{a}{4}+b-\frac{3a}{4}\gamma_5 \right) d \nonumber \\
&+ & \frac{g'}{3}\bar{u}\slashed{Z'} \left(\frac{5a}{4}+b+\frac{3a}{4}\gamma_5 \right)u \nonumber \\
& + & g'\bar{e}\slashed{Z'}\left(-\frac{3a}{4}-b-\frac{a}{4}\gamma_5\right)e.
\end{eqnarray}
In Fig.~\ref{annihilation}, we show the expected annihilation cross section into $\gamma \gamma $ and $Z\gamma$ for the case $a=1$ and $b=-5/4$, motivated by the embedding in GUTs; the Abelian gauge symmetry with $a=1$ and $b=-5/4$ is obtained from $SO(10)$. We show the numerical predictions for different choices of the new gauge boson mass together with the experimental bounds from Fermi-LAT~\cite{Ackermann:2015lka,Ackermann:2013uma} and H.E.S.S.~\cite{Abramowski:2013ax}  collaborations. We note that these DM annihilation processes are independent of the parameter $b$. 
For the predictions shown in Fig.~\ref{annihilation}, a relic abundance of $\Omega h^2 = 0.12$ is assumed and we do not stick to any specific mechanism to explain the relic density.
\begin{figure}[h]
\centering
\includegraphics[width=0.95\linewidth]{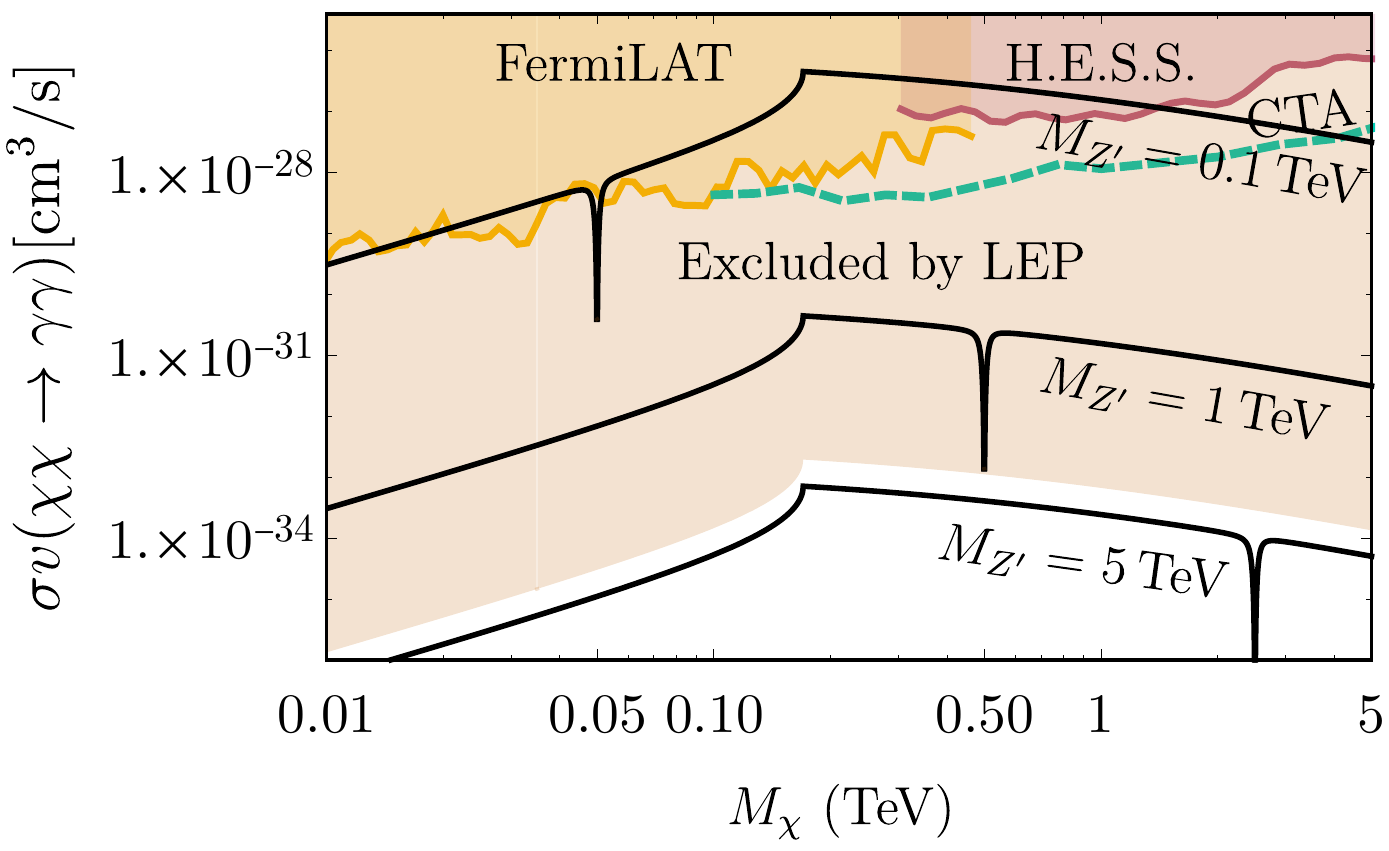} 
\includegraphics[width=0.95\linewidth]{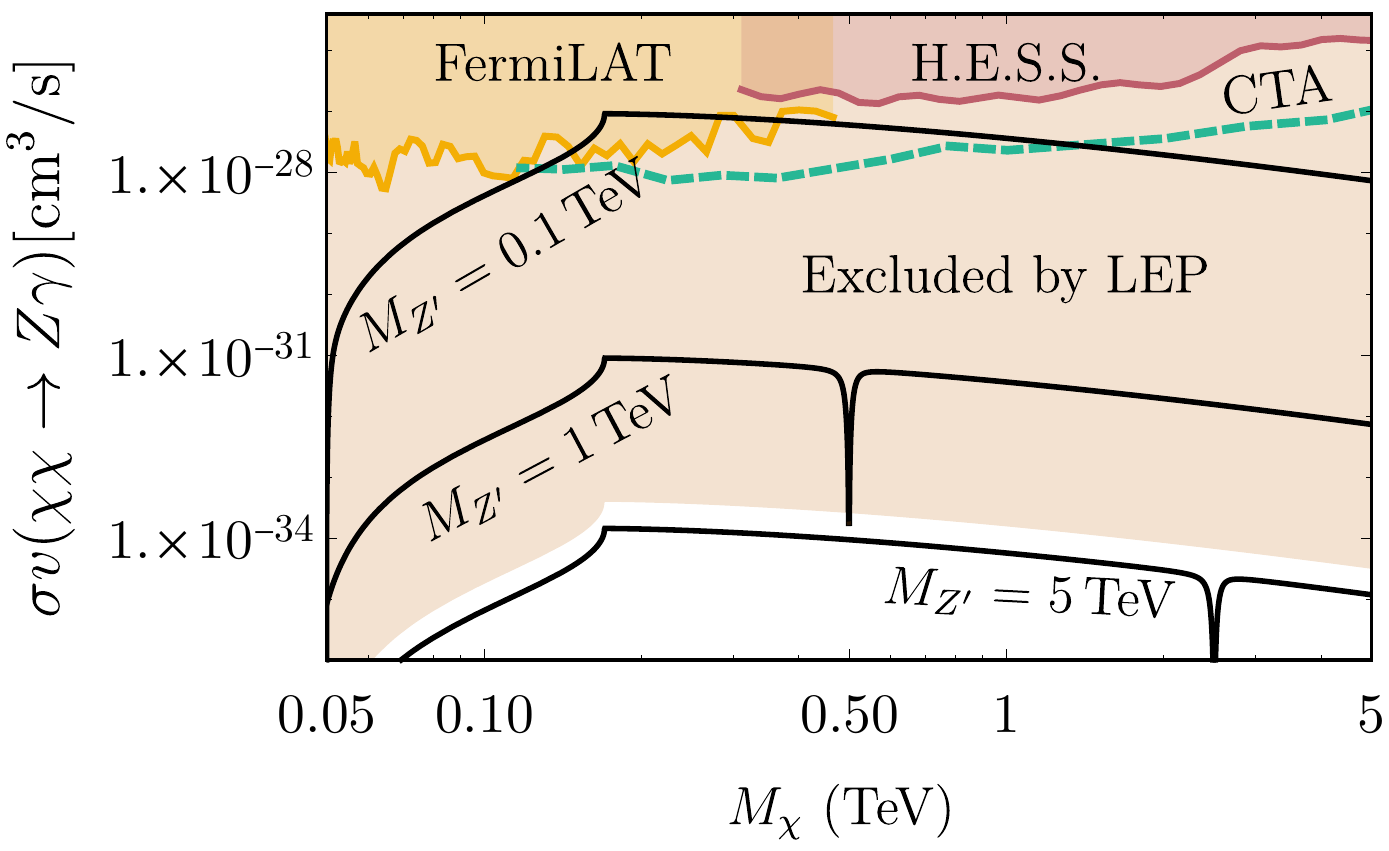} 
\caption{Annihilation cross sections for $\chi \chi \to \gamma \gamma$ (upper panel) and $\chi \chi \to Z \gamma$ (lower panel) in Model I when $a=1$, $b=-5/4$ and $g'=1$. We show in yellow and purple the excluded areas by the Fermi-LAT~\cite{Ackermann:2015lka,Ackermann:2013uma} and H.E.S.S~\cite{Abramowski:2013ax} collaborations, respectively. The shaded brown area is excluded by the LEP2~\cite{Alioli:2017nzr} bounds. The projected bounds by the CTA collaboration~\cite{Acharya:2017ttl} are shown by the dashed green line.
Here, we use the value $J_\text{ann}=13.9 \times 10^{22} \text{ GeV}^2 \text{cm}^{-5}$~\cite{Ackermann:2015lka,Ackermann:2013uma} for our numerical analysis.}
\label{annihilation}
\label{diagrams}
\end{figure}

As the figure~\ref{annihilation} shows, for light masses of the new mediator, one can hope to test these signals. However, since the new mediator couples to the charged leptons, the model is sensitive to the collider bounds, particularly to the strong bound coming form LEP2~\cite{Alioli:2017nzr}. Unfortunately, the LEP2 bounds, together with some other bounds coming from dijet searches at the LHC~\cite{FileviezPerez:2018jmr} forbid the parameter space where the predictions are close to the experimental 
bounds, see the shaded brown area in Fig.~\ref{annihilation}. Therefore, only if the experimental bounds are improved by several orders of magnitude in the future, one could hope to test this model.

\newpage

\item {\bf{Model II}}: Let us consider a simple model where the new gauge boson is leptophobic with very suppressed couplings to the SM leptons. 
Let us assume that the new force couples only to the quarks as follows:
\begin{eqnarray*}
q_L &\sim& (3,2,1/6,n_L), \\
 u_R &\sim& (3,1,2/3,n_R),\\
 d_R &\sim& (3,1,-1/3,n_R).
\end{eqnarray*}
The relevant interactions for our discussion are given by
\begin{eqnarray}
{\cal L} & \supset & g' \bar{u} \slashed{Z^{'}} \left ( \frac{n_L + n_R}{2}+ \frac{n_R - n_L}{2} \gamma_5 \right) u   \nonumber \\
 & +& g' \bar{d} \slashed{Z^{'}} \left ( \frac{n_L + n_R}{2} + \frac{n_R - n_L}{2} \gamma_5 \right) d.
\end{eqnarray}
This case is similar to the one presented in the last section with the difference that, now, the new gauge boson does not couple to the leptons and, therefore, the collider bounds are weaker than in the previous theory. Therefore, one could have large cross sections for the gamma lines in agreement with the experiment. See Ref.~\cite{FileviezPerez:2018jmr} for a detailed discussion of the collider bounds for leptophobic gauge bosons. Fig.~2 of the cited reference shows the bounds coming from direct searches at the LHC on a leptophobic gauge coupling. As one can appreciate, they allow for interesting regions of the parameter space in which the ratio $g' / M_{Z'}$ is not as constrained as the LEP bound ($g'/M_{Z'} < 7 $ TeV). We would like to mention that this model can be anomaly free if we add vector-like fermions to cancel all anomalies. Here, we are not interested in the details of a specific model.
\vspace{0.2cm}
\\
\begin{figure}[h]
\centering
\includegraphics[width=0.95\linewidth]{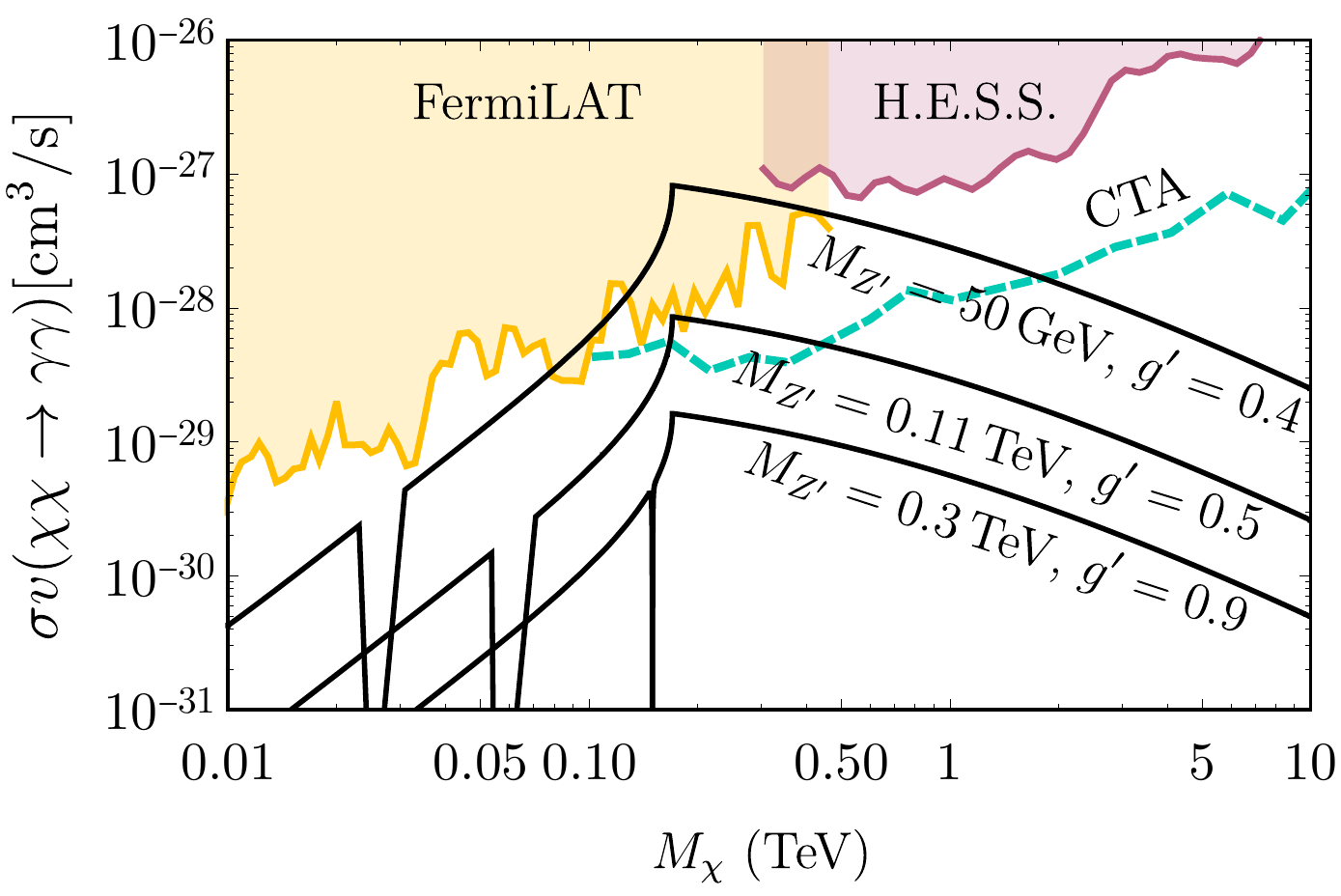} 
\includegraphics[width=0.95\linewidth]{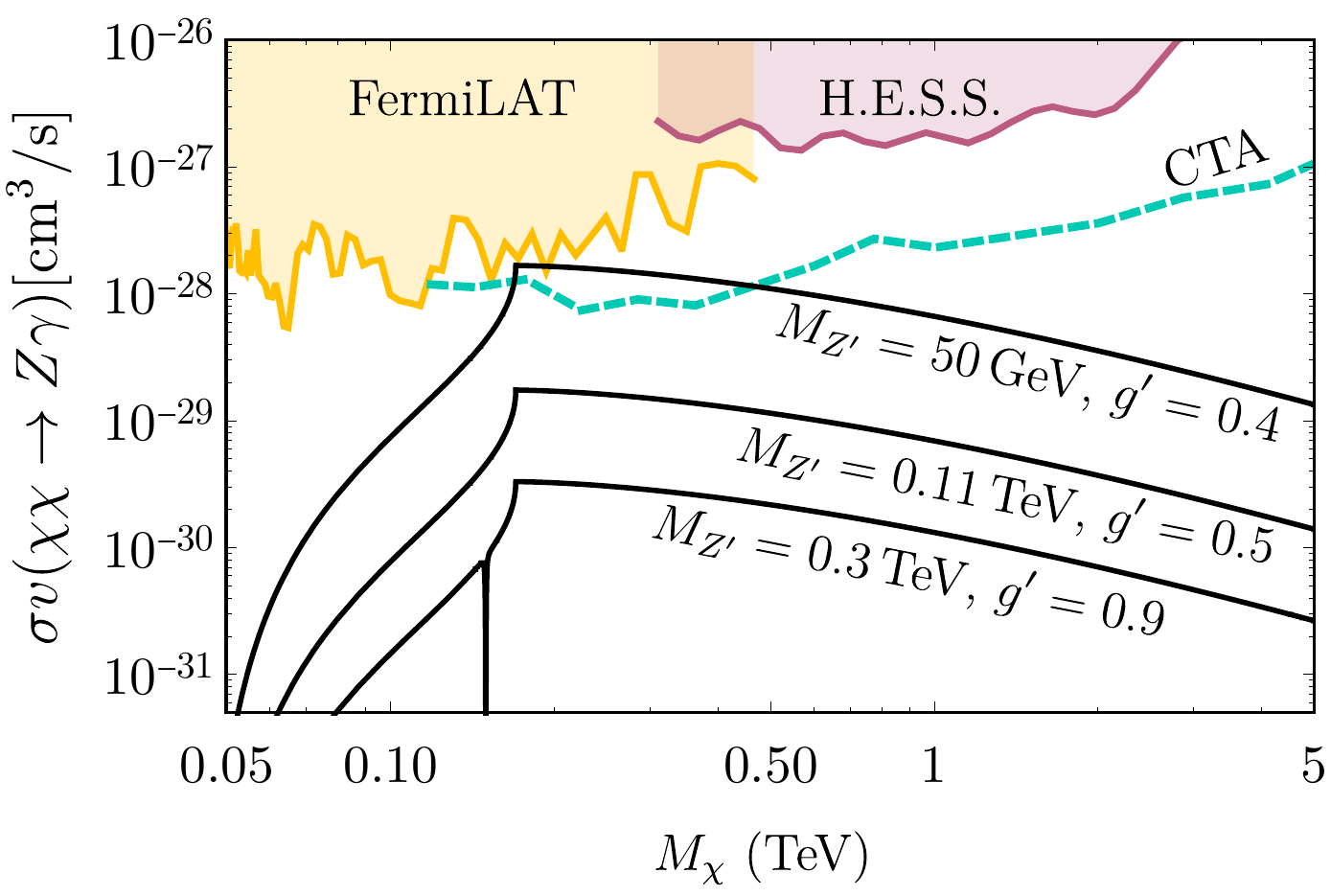} 
\caption{Cross section for the dark matter annihilation into two photons (upper panel) and a photon and Z boson (lower panel) in Model II. Here, we have chosen $n_L=0$ and $n_R=1/3$ as a representative charge 
and different values for the $Z'$ mass. We show in yellow and purple the excluded areas by the Fermi-LAT~\cite{Ackermann:2015lka,Ackermann:2013uma} and H.E.S.S~\cite{Abramowski:2013ax} collaborations, respectively. The projected bounds by the CTA collaboration~\cite{Acharya:2017ttl} are shown by the dashed green line.
Here, we use the value $J_\text{ann}=13.9 \times 10^{22} \text{ GeV}^2 \text{cm}^{-5}$~\cite{Ackermann:2015lka,Ackermann:2013uma} for our numerical analysis.}
\label{annihilationLA}
\label{diagrams}
\end{figure}
In Fig.~\ref{annihilationLA}, we show the two relevant dark matter annihilation cross sections for the emission of gamma lines. Here, we show the predictions with the SM fermions, since are the ones that will contribute the most in the annihilation cross sections through the fermionic loop, and assume that, given $n_L$ and $n_R$, the model has the right UV-fermions to satisfy anomaly cancellation. This assumption is reflected in the use of Eq.~\ref{ACgg} and Eq.~\ref{ACgZ} for the cross sections. 
We can see, in Fig.~\ref{annihilationLA}, that the cross sections for gamma lines can be large in agreement with all experimental constraints. Therefore, there is a hope to test these predictions in current or future experiments. We note that for the charge assignment assumed in Fig.~\ref{annihilationLA}, $n_L= 0$ and $n_R=1/3$, the vector coupling of the leptophobic mediator with quarks is $g'/6$ and one needs to take into account that $g_B = g'/2$ when applying the experimental bounds for a leptophobic gauge boson shown in Fig.~2 from Ref.~\cite{FileviezPerez:2018jmr}.
See Ref.~\cite{Duerr:2015vna} for the predictions of the gamma lines in a model where extra fermions in the loop can generate a large cross section for the gamma lines.
\end{itemize}
%%%%%%%%%%%%%%%%%%%%
\section{Gamma Lines Spectrum}
%%%%%%%%%%%%%%%%%%%%
%%%%%%%%%%%%%%%%%%%%%%%%%%%%
\begin{figure}[t]
\centering
\includegraphics[width=0.9\linewidth]{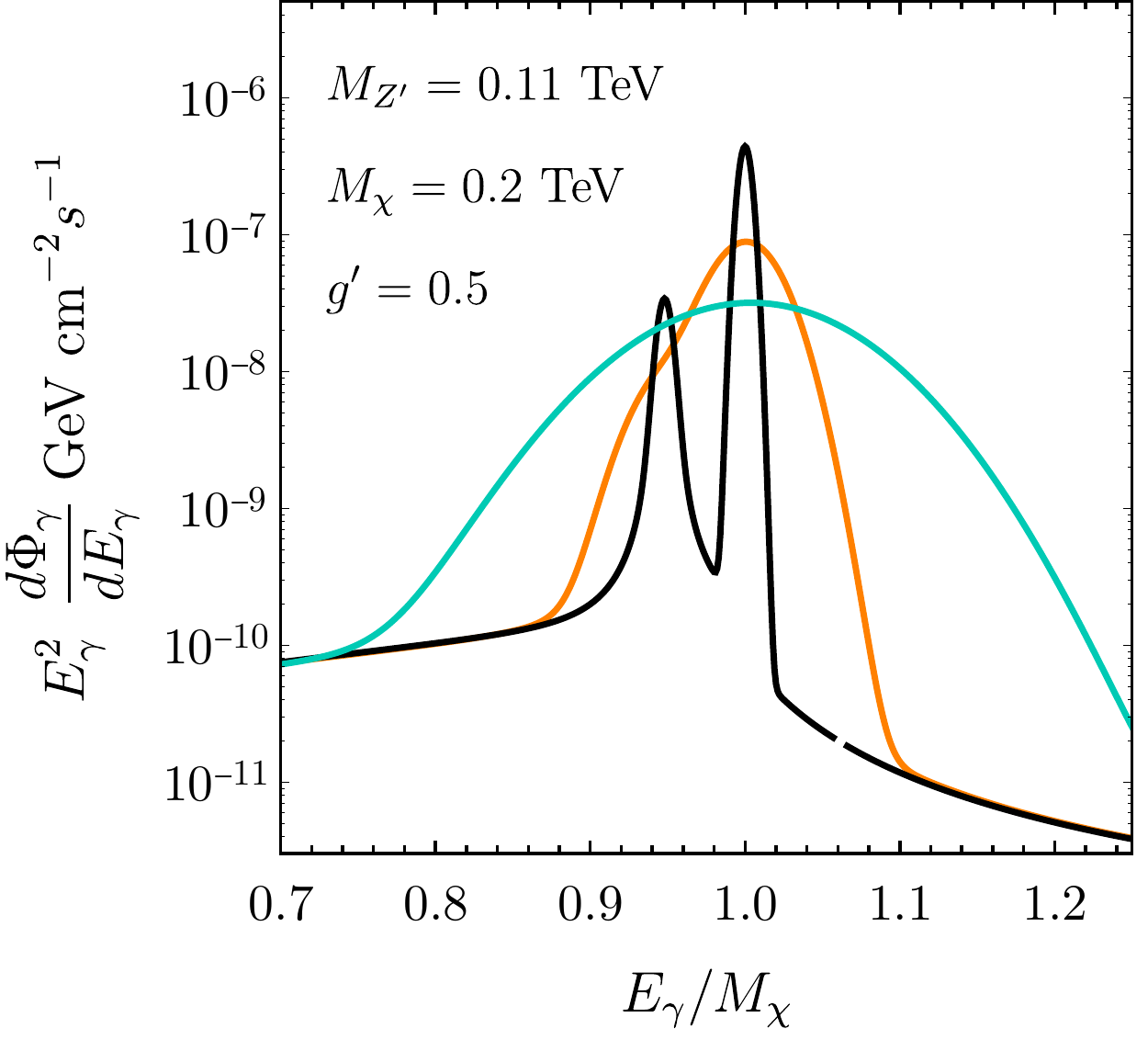} 
\includegraphics[width=0.9\linewidth]{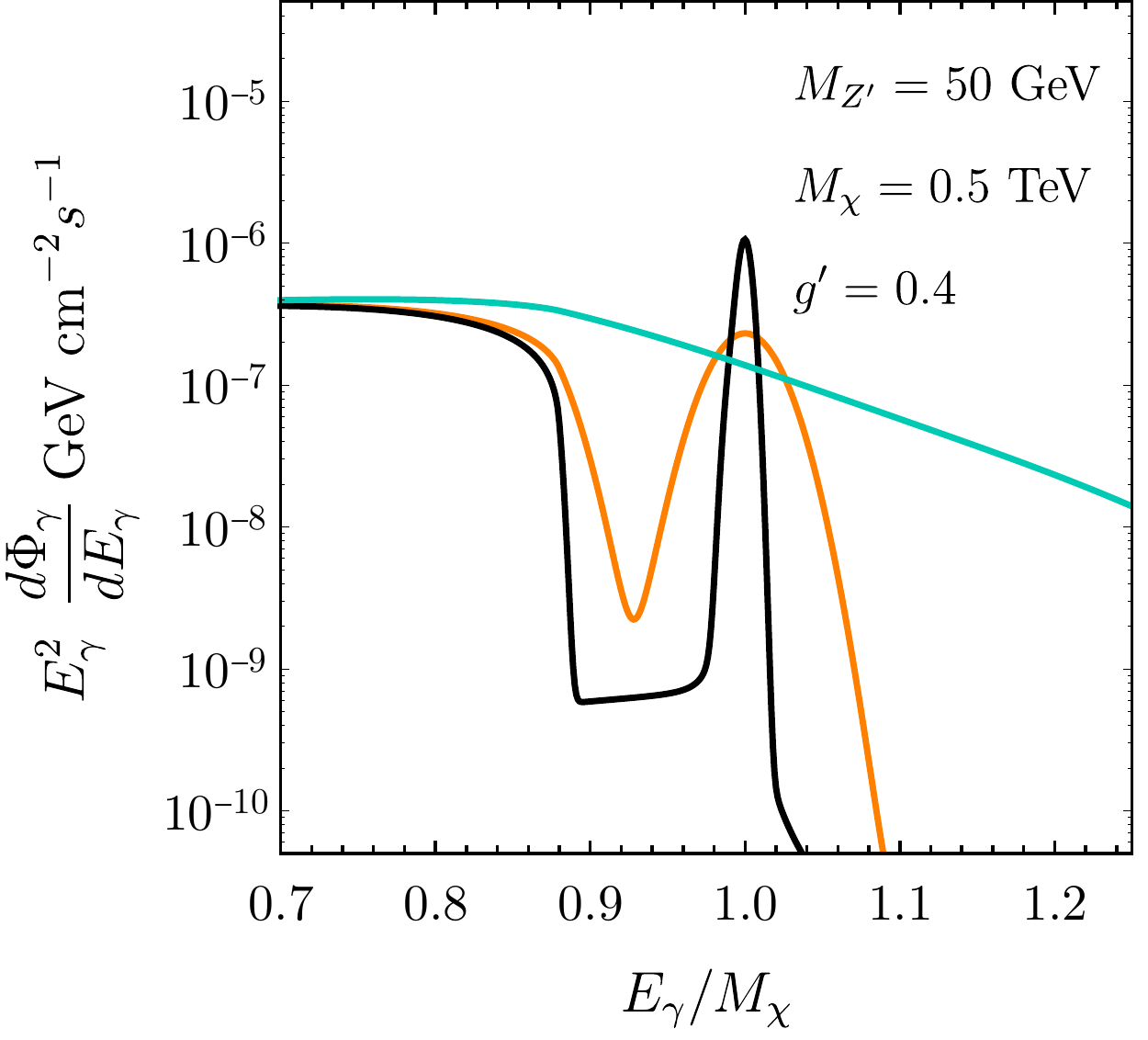} 
 \includegraphics[width=0.7\linewidth]{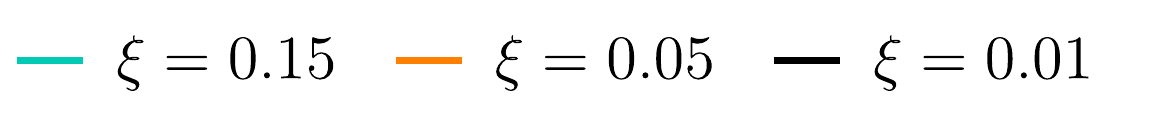}
\caption{Differential spectrum of the annihilation of DM into gamma rays ($\gamma \gamma$ and $Z\gamma$) in the context of Model II, assuming an energy resolution of $\xi= 0.01$ in black, $\xi = 0.05$ in orange, 
and $\xi = 0.15$ in turquoise. Two different scenarios are shown: in the lower panel, $M_\chi = 0.5$ TeV, $M_{Z'}=50$ GeV and $g'=0.4$; in the upper panel, $M_\chi = 0.2$ TeV, $M_{Z'}=0.11$ TeV and $g' = 0.5$. Here, we have chosen $n_L=0$ and $n_R=1/3$ as a representative charge 
and different values for the $Z'$ mass. Here, we use the value $J_\text{ann}=13.9 \times 10^{22} \text{ GeV}^2 \text{cm}^{-5}$~\cite{Ackermann:2015lka,Ackermann:2013uma} for our numerical analysis. }
\label{gammalines-spectrum}
\end{figure}
%%%%%%%%%%%%%%%%%%%%%%%%%%%%
We have discussed two simple classes of models for Majorana dark matter where the final state radiation processes are highly suppressed and one can have large cross sections for the gamma lines.
Here, we show the spectrum for these gamma lines. The flux for the gamma lines is given by
\begin{equation}
\frac{d\Phi_\gamma}{dE_\gamma}= \frac{n_\gamma }{8 \pi M_\chi^2} \frac{ d\langle \sigma v_\text{rel} \rangle}{dE_\gamma}J_\text{ann}
=\frac{n_\gamma \langle \sigma v_\text{rel} \rangle}{8\pi M_\chi^2}\frac{dN}{dE_\gamma} J_\text{ann},
\end{equation}
where, in the last equality, the narrow width approximation has been applied. The J-factor, $J_\text{ann}$, encodes all astrophysical assumptions made regarding the dark matter distribution. 
Here, we will use the value $J_\text{ann}=13.9 \times 10^{22} \text{ GeV}^2 \text{cm}^{-5}$~\cite{Ackermann:2015lka,Ackermann:2013uma} for our numerical analysis, which is the J-factor used by Fermi-LAT for the R3 region-of-interest. 
The spectrum function is given by
\begin{equation}
\frac{dN}{dE_\gamma}=\int_0^\infty dE_0 \,  W_\text{final} \, G(E_\gamma, \xi /\omega, E_0),
\end{equation}
where $W_\text{final}$ is $W_{\gamma \gamma} = \delta(E_0 - M_\chi)$ for the annihilation into two photons and
\begin{equation}
W_{\gamma Z} = \frac{1}{\pi}\frac{4M_\chi M_Z \Gamma_Z}{(4M_\chi^2-4M_\chi E_0 - M_Z^2)^2+ \Gamma_{Z}^2M_Z^2},
\label{BWS}
\end{equation}
for the $Z \gamma$ line. Here, we use a Gaussian function to model the detector resolution, $G(E_\gamma,\xi/\omega, E_0)$, which reads as
\begin{equation}
G(E_\gamma,\xi/\omega, E_0)=\frac{1}{\sqrt{2\pi}E_0(\xi/\omega)} e^{ -\frac{(E_\gamma - E_0)^2}{2E_0^2(\xi/\omega)^2}},
\end{equation}
where $\xi$ is the energy resolution and  $\omega = 2 \sqrt{2 \rm{log} 2} \approx 2.35$ determines the full width at half maximum, with the standard deviation given by $\sigma_0=E_0 \xi/w$.

In Fig.~\ref{gammalines-spectrum}, we show the predictions for the spectrum for the gamma lines $\gamma \gamma$ and $Z \gamma$ for the second model discussed above, where the new gauge boson couples only to the SM quarks. In this figure, different energy resolutions are assumed: $\xi = 0.01$ in black, $\xi = 0.05$ in orange and $\xi = 0.15$ in turquoise. Here, we assume $n_R=1/3$ and $n_L=0$ for illustration. We show two different scenarios: 

In the upper panel, we show the spectrum for the gamma lines when $M_{Z^{'}}=110$ GeV, $M_\chi=200$ GeV, and $g^\prime=0.5$. In this figure, one can clearly see the two gamma lines $\gamma \gamma$ and $Z \gamma$ if the energy resolution is good enough. We note that for the annihilation into two photons, the energy of the gamma line is at $E_\gamma = M_\chi$, whereas the energy of the $Z \gamma$ line is given by Eq.~(\ref{Eq1}). These two energies correspond to the peaks shown in the aforementioned plot. Here, the FSR processes are velocity and helicity suppressed in the region of energies shown in the figure.

On the other hand, in the lower panel we show the spectra for $M_{\chi}=500$ GeV, $M_{Z^{'}}=50$ GeV and $g^\prime=0.4$. As one can see, here the FSR strongly contributes in the left-hand-side of the figure, unfortunately making it impossible to distinguish the gamma line from the background for the $15\%$ energy resolution. We note that, in this case, $M_{Z'} < M_t$, and therefore the FSR processes are not helicity suppressed anymore (see Eq.~(\ref{Eq3})). As the figure shows, after the maximal energy of the photon in the $\chi \chi \to \bar{t} t \gamma$ processes, given by Eq.~(\ref{Eq2}), the continuum suffers a suppression of three orders of magnitude, which allows us to observe the peak of the gamma lines. Notice that, for this scenario, the ratio $M_{Z'}^2/M_\chi^2$ is very small and, therefore, the two peaks $\gamma \gamma$ and $Z \gamma$ are located approximately at $M_\chi$ and cannot be easily resolved. We also note that the continuum that extends along the right-hand-side of the $Z \gamma$ peaks is a consequence of the Breit-Wigner like shape from Eq.~\ref{BWS}, which can only be distinguished from the gaussian used to model the detector resolution in the scenarios with higher resolution.

We remark that, in these scenarios, the contribution from the final state radiation to the continuum spectrum is highly suppressed, which is crucial for the gamma lines to be distinguished from the continuum spectrum. These striking results tell us that one can have simple gauge theories for dark matter where the gamma lines form dark matter annihilation can be visible and the predictions can be close to the current experimental limits by Femi-LAT and H.E.S.S. collaborations.

%%%%%%%%%%%%%%%%
\section{Summary}
%%%%%%%%%%%%%%%%
There are many ways to look for signatures in direct and indirect dark matter experiments, but gamma lines from dark matter 
annihilation are, perhaps, the most striking signatures which can reveal the nature of dark matter. 
In this article, we have discussed the visibility of gamma lines from dark matter annihilation in simple theories. 
We have shown that, in theories where the dark matter candidate is a Majorana fermion, the final state radiation 
processes are velocity suppressed and the gamma line can be seen above the continuum spectrum.
We have studied two main classes of theories where the mediator between the Hidden and Standard Model sectors 
is a new Abelian gauge boson. In this context, we show that for a gamma line to be clearly detected (a) the dark matter must be Majorana, and (b) at least one fermion in the theory must couple to the new mediator in an axial way.

We investigated two main classes of theories where the gamma lines can be distinguished 
from the continuum spectrum. In the first case, the new gauge boson couples to all standard model fermions and, although the gamma lines can be visible, the collider bounds rule out the parameter space where the cross section for gamma lines can be large. In the second class of theories, the gauge boson does not couple to leptons, and one can have large values for the annihilation cross sections for gamma lines in agreement with all collider bounds. These interesting results could help us to identify the theory for dark matter at the current or future experiments. 

{\textit{Acknowledgments:}} We would like to thank Mark B. Wise for discussions and P. F. P. thanks the Walter Burke Institute for Theoretical Physics at Caltech for hospitality. The work of C.M. has been supported in part by Grants No. FPA2014-53631-C2-1-P, FPA2017-84445-P and SEV-2014-0398 (AEI/ERDF, EU), and La Caixa-Severo Ochoa scholarship. C.M. also thanks Case Western Reserve University for hospitality.  We thank the referees for the very useful comments.

\appendix

\begin{widetext}

%%%%%%%%%%%%%%%%%%%%%%%%%%%
\section{Cross Sections for the Gamma Lines}
%%%%%%%%%%%%%%%%%%%%%%%%%%%
The annihilation cross section for the different annihilation channels mediated by the new gauge boson $Z'$ are given by:
\begin{itemize}
\item Cross section for the ${\chi}\chi \to \gamma \gamma$ annihilation:
\begin{eqnarray}
 \sigma v ({{\chi}\chi \to \gamma \gamma}) &=&  \frac{\alpha^2}{4\pi^3} \frac{{(g')}^4 n_\chi^2 M_\chi^2}{M_{Z'}^4}\frac{(4M_\chi^2 - M_{Z'}^2)^2}{(4M_\chi^2-M_{Z'}^2)^2+\Gamma_{Z'}^2M_{Z'}^2} \left | \sum_f N_c^f n_A^f Q_f^2(2M_f^2 C_0^\gamma +1)\right |^2, \nonumber \\
&\stackrel{A.C.}{=}&  \frac{\alpha^2}{\pi^3} \frac{{(g')}^4 n_\chi^2 M_\chi^2}{M_{Z'}^4}\frac{(4M_\chi^2 - M_{Z'}^2)^2}{(4M_\chi^2-M_{Z'}^2)^2+\Gamma_{Z'}^2M_{Z'}^2} \left | \sum_f N_c^f n_A^f Q_f^2M_f^2 C_0^\gamma \right |^2. 
\label{ACgg}
\end{eqnarray}
\item Cross section for the ${\chi}\chi \to \gamma Z$ annihilation: 
\begin{eqnarray}
\sigma v (\chi \chi \to \gamma Z) &=& \frac{\alpha \, (g')^4 n_\chi^2 }{512 \pi^4}\frac{(4M_\chi^2-M_Z^2)^3}{ (4M_\chi^2-M_{Z'}^2)^2+\Gamma_{Z'}^2 M_{Z'}^2}\frac{(M_{Z'}^2-4M_\chi^2)^2}{M_\chi^4M_{Z'}^4} \left| \sum_f N_c^f Q_f \left(n_V^f g_{A}^f + n_A^f g_{V}^f  (2M_f^2 C_0^Z+1)\right) \right|^2 \nonumber \\
&\stackrel{A.C.}{=}&\frac{\alpha \, (g')^4 n_\chi^2 }{512 \pi^4}\frac{(4M_\chi^2-M_Z^2)^3}{ (4M_\chi^2-M_{Z'}^2)^2+\Gamma_{Z'}^2 M_{Z'}^2}\frac{(M_{Z'}^2-4M_\chi^2)^2}{M_\chi^4M_{Z'}^4} \left| \sum_f N_c^f Q_f  n_A^f g_{V}^f 2M_f^2 C_0^Z \right|^2.
\label{ACgZ}
\end{eqnarray}
\item Cross section for the ${\chi}\chi \to \gamma h$ annihilation:
\begin{eqnarray}
&&  \sigma v ({\chi}\chi \to \gamma h) = \nonumber \\
&& v^2 \frac{{(g')}^4n_\chi^2 \alpha}{768\pi^4 M_\chi^2}\frac{4M_\chi^2-M_h^2}{ (4M_\chi^2-M_{Z'}^2)^2+\Gamma_{Z'}^2M_{Z'}^2}
 \left| \sum_f  N_c^f g_S^f \, n_{V}^f M_f Q_f \left( \frac{8 M_\chi^2}{4M_\chi^2-M_h^2}(\Lambda^f_\chi -\Lambda^f_h) +2+(4M_\chi^2-M_h^2+4M_f^2)C_0^h \right)\right|^2 \nonumber \\
\end{eqnarray}
\end{itemize}

where we have the following loop functions:
\begin{eqnarray}
C_0^A &=& C_0(0,M_A^2,s; M_f,M_f,M_f),\\
\Lambda^f_\chi &=&\Lambda(4M_\chi^2;M_f,M_f),\\
\Lambda^f_h &=& \Lambda(M_h^2;M_f,M_f).
\end{eqnarray}
and we have applied the anomaly cancellation conditions when simplifying the expressions. See Ref.~\cite{Duerr:2015wfa} for the calculation 
of the gamma lines cross section in gauge theories for dark matter.
%%%%%%%%%%%%%%%%%
\section{Final State Radiation}
%%%%%%%%%%%%%%%%%
%
The amplitude squared of the processes contributing to the final state radiation can be written as an expansion on the velocity, 
\begin{equation}
|{\cal M}|^2_\text{FSR} = \frac{M_f^2}{M^2_{Z'}}A +  v^2 B + {\cal O}(v^4),
\label{EqAp}
\end{equation}
where the leading order term of the expansion is suppressed by the ratio $(M_f/M_Z^\prime)^2$. The coefficients that parametrise the above expression read as,
\begin{eqnarray}
&&A=16 \pi \, \alpha \, {(g^\prime)}^4  n_\chi^2 Q_f^2 N_c^f  (M_{Z^\prime}^2-4M_\chi^2)^2\times  \\
&&\frac{{(n_A^f)}^2(E_f-M_\chi)^2\left(2(E_f-M_\chi)(E_f+E_\gamma-M_\chi)+M_f^2\right) + {(n_V^f)}^2(E_f+E_\gamma - M_\chi)^2\left(2(E_f-M_\chi)(E_f + E_\gamma -M_\chi)-3M_f^2 \right) }{M_{Z^\prime}^2(E_f-M_\chi)^2(E_f+E_\gamma-M_\chi)^2((4M_\chi^2-M_{Z^\prime}^2)^2+\Gamma_{Z^\prime}^2M_{Z^\prime}^2)} \nonumber  \\
&&B=16 \pi \, \alpha \,  {(g^{\prime})}^4 m_\chi^2 n_\chi^2 Q_f^2 N_c^f ({(n_A^f)}^2+{(n_V^f)}^2) \times\\
&&  \frac{ \left(2E_f M_\chi (E_\gamma^2-3 E_\gamma M_\chi + 2 M_\chi^2)-2 E_f^4 - 2 E_f^3 ( E_\gamma - 2 M_\chi) -E_f^2(E_\gamma^2-6E_\gamma M_\chi + 6M_\chi^2)-2M_\chi^2(E_\gamma - M_\chi)^2\right)}
{M_{Z^\prime}^2 (E_f+E_\gamma-M_\chi)^2((4M_\chi^2-M_{Z^\prime}^2)^2+\Gamma_{Z^\prime}^2M_{Z^\prime}^2)}, \nonumber 
\end{eqnarray}
The DM annihilation to $\gamma \bar{f}f$ is described by a three-body phase space and, therefore, the cross section is given by
\begin{equation}
\frac{d^2 \langle \sigma v_\text{rel} \rangle}{dE_\gamma dE_f}=\frac{1}{32\pi^3 s}|{\cal M}|^2_\text{FSR} \, \theta(\cos \theta_{f\gamma}^2-1),
\end{equation}
In order to compute the continuum spectrum of the FSR, given by 
\begin{equation}
\frac{d\Phi_\gamma}{dE_\gamma}= \frac{n_\gamma }{8 \pi M_\chi^2} \frac{ d\langle \sigma v_\text{rel} \rangle}{dE_\gamma}J_\text{ann},
\end{equation}
one needs to integrate the differential cross section with respect to $E_f$, which kinematic range is determined by the condition $\cos \theta_{1\gamma}^2 \leq 1 $. According to this, the integration limits are given by 
\begin{equation}
E_f^\pm = M_\chi - \frac{E_\gamma}{2} \pm \frac{E_\gamma}{2}\sqrt{1+\frac{M_f^2}{(E_\gamma - M_\chi)M_\chi}}.
\end{equation}

\FloatBarrier
%%%%%%%%%%%%%%%%%
\section{Experimental Bounds}
%%%%%%%%%%%%%%%%%
Here we list the bounds from the Fermi-LAT~\cite{Ackermann:2015lka,Ackermann:2013uma} and H.E.S.S~\cite{Abramowski:2013ax} collaborations for the DM annihilation into two gamma rays, $\rm{DM} \ \rm{DM} \to \gamma \gamma$, and the corresponding bounds for the annihilation $\rm{DM} \ \rm{DM} \to Z \gamma$. See the table below for the upper bounds on these cross sections.

\begin{table}[h]
\begin{tabular}{ c  c  c  c  | c c c c }
\multicolumn{4}{c}{Fermi-LAT collaboration} & \multicolumn{4}{c}{H.E.S.S. collaboration}\\
$E_\gamma$ [GeV] & $ \sigma v (\gamma \gamma)[ 10^{-29} \text{cm}^{3}/s]$ & $M_\chi$ [GeV] & $\sigma v (Z \gamma) [ 10^{-29} \text{cm}^{3}/s]$ & $E_\gamma$ [TeV] & $\sigma v (\gamma \gamma) [10^{-27} \text{cm}^{3}/s$] &  $M_\chi$ [TeV] & $\sigma v (Z \gamma) [10^{-27} \text{cm}^{3}/s]$ \\
\hline
\hline
20.0		&	2.00		&	56.7		&	32.1  	& 	0.304	&	1.09		&	0.311		&	 2.28\\
21.1		&	0.946	&	57.4		&	14.0 	 	&	0.333	&	0.845	&	0.339		&	1.75 \\
22.3		&	0.946	&	58.1		&	12.8 	 	&	0.364	&	0.786	&	0.370		&	1.62 \\
23.6		&	0.957	&	58.9		&	11.9 	 	&	0.396	&	0.936	&	0.401		&	1.92	\\
24.9		&	0.829	&	59.7		&	9.54  	&	0.436	&	1.12		&	0.441		&	2.28	\\
26.4		&	0.888	&	60.7		&	9.38 		&	0.474	&	0.985	&	0.478		&	2.01 \\
27.9		&	1.24		&	61.6		&	12.1 		&	0.518	&	0.694	&	0.522		&	1.41 \\
29.5		&	0.958	&	62.7		&	8.65		&	0.567	&	0.665	&	0.571		&	1.35 \\
31.2		&	0.661	&	63.8		&	5.53 		&	0.620	&	0.861	&	0.623		&	1.74 \\
33.0		&	0.695	&	65.0		&	5.39		&	0.678	&	0.916	&	0.681		&	1.85 \\
34.9		&	1.42		&	66.3		&	10.2 		&	0.742	&	0.785	&	0.745		&	1.58 \\
36.9		&	3.08		&	67.6		&	20.7		&	0.811	&	0.730	&	0.814		&	1.47 \\
39.0		&	3.87		&	69.1		&	24.3 		&	0.881	&	0.817	&	0.884		&	1.64 \\
41.3		&	3.79		&	70.7		&	22.2		&	0.964	&	0.925	&	0.966		&	1.86 \\
43.8		&	6.41		&	72.5		&	35.1		&	1.05		&	0.843	&	1.06		&	1.69 \\
46.4		&	6.54		&	74.4		&	33.6 		&	1.15		&	0.768	&	1.16		&	1.54	\\
49.1		&	5.62		&	76.3		&	27.2 		&	1.27		&	0.906	&	1.27		&	1.82	\\
52.1		&	3.12		&	78.6		&	14.2 		&	1.38		&	1.11		&	1.38		&	2.23\\
55.2		&	3.38		&	80.9		&	14.5 		&	1.52		&	1.37		&	1.52		&	2.74\\
58.6		&	7.13		&	83.5		&	29.0 		&	1.65		&	1.49		&	1.65		&	2.98\\
62.2		&	6.95		&	86.3		&	26.8 		&	1.79		&	1.37		&	1.79		&	2.74\\
66.0		&	4.59		&	89.3		&	16.8 		&	1.98		&	1.29		&	1.98		&	2.57\\
70.1		&	5.18		&	92.6		&	18.1		&	2.15		&	1.44		&	2.15		&	2.88 \\
74.5		&	5.56		&	96.1		&	18.5 		&	2.35		&	2.00		&	2.35		&	4.01\\
79.2		&	3.08		&	100		&	9.82		&	2.57		&	3.16		&	2.57		&	6.31\\
84.2		&	2.87		&	104		&	8.79 		&	2.81		&	4.97		&	2.81		&	9.94\\
89.6		&	2.87		&	109		&	8.45		&	3.07		&	5.98		&	3.07		&	12.0 \\
95.4		&	2.82		&	114		&	8.01 		&	3.36		&	5.74		& 	3.36		&	11.5 \\
102		&	5.77		&	119		&	15.8 		&	3.67		&	6.17		&	3.67		&	12.3\\
108		&	5.73		&	125		&	15.3 		&	4.01		&	7.58		&	4.01		&	15.2\\
115		&	15.2		&	131		&	39.4		&	4.36		&	7.90		&	4.36		&	15.8 \\
123		&	15.1		&	138		&	38.1 		&	4.77		&	7.42		&	4.77		&	14.8\\
131		&	10.8		&	145		&	26.6 		&	5.22		&	7.27		&	5.22		&	14.5	\\
140		&	5.29		&	154		&	12.7 		&	5.71		&	7.19		&	5.71		&	14.4	\\
150		&	10.6		&	163		&	25.0		&	6.24		&	6.69		&	6.24		&	13.4 \\
160		&	8.15		&	172		&	18.9 		&	6.83		&	7.97		&	6.83		&	16.0	\\
171		&	13.0		&	182		&	29.6 		&	7.47		&	11.3		&	7.47		&	22.7	\\
183		&	6.68		&	194		&	15.0 		&	8.16		&	15.9		&	8.16		&	31.8	\\
196		&	13.3		&	206		&	29.4		&	8.87		&	19.2		&	8.87		&	38.3	 \\
210		&	9.19		&	220		&	20.1 		&	9.70		&	17.1		&	9.70		&	34.2	\\
225		&	13.0		&	234		&	28.1 		&	10.6		&	12.2		&	10.6	&	24.3	\\
241		&	18.7		&	249		&	40.0 		&	11.6		&	11.0		&	11.6	&	21.9	\\
259		&	10.2		&	267		&	21.7 		&	12.7		&	14.7		&	12.7	&	29.3	\\
276		&	41.2		&	283		&	86.8 		&	13.9		&	21.0		&	13.9	&	42.0	\\
294		&	41.4		&	301		&	86.7		&	15.2		&	27.8		&	15.2	&	55.5 \\
321		&	17.3		&	327		&	36.0 		&	16.6		&	36.7		&	16.6	&	73.4	\\
345		&	15.0		&	351		&	31.0 		&	18.2		&	41.1		&	18.2	&	82.2	\\
367		&	48.7		&	373		&	100 		&	19.7		&	39.4		&	19.7	&	78.9	\\
396		&	51.7		&	401		&	106 		&	21.7		&	39.4		&	21.7	&	78.8 \\
427		&	49.6		&	432		&	101  		&	23.6		&	49.5		&	23.6	&	98.9	\\
462		&	39.4		&	467		&	80.3 		&	25.8		&	81.2		&	25.8	&	162	\\

\end{tabular}
\end{table}

\end{widetext}

\FloatBarrier

%%%%%%%%%%%%%%%%%%%%%%%%%%%%%%%%%%%%%%%%%%%%%%%%%%%%%%%%%%%%%%%%%%%%%%%%%%%%%%%

%%%%%%%%%%%%%%%%%%%%%%%%%%%%%%%%%%%%%%%%%%%%%%%%%%%%%%%%%%%%%%%%%%%%%%%%%%%%%%%


\begin{thebibliography}{99}
%%%%%%%%%%%%%%%%%%%%%%%%%%%%%%%%%%%%%%%%%%%%%%%%%%%%%%%%%%%%%%%%%%%%%%%%%%%%%%%


\bibitem{CMB}
M. Tanabashi et al. (Particle Data Group), Phys. Rev. D 98, 030001. 


%\cite{Bringmann:2012ez}
\bibitem{Bringmann:2012ez}
  T.~Bringmann and C.~Weniger,
  ``Gamma Ray Signals from Dark Matter: Concepts, Status and Prospects,''
  Phys.\ Dark Univ.\  {\bf 1} (2012) 194,
%  doi:10.1016/j.dark.2012.10.005
  [arXiv:1208.5481 [hep-ph]].
  %%CITATION = doi:10.1016/j.dark.2012.10.005;%%
  %197 citations counted in INSPIRE as of 21 Jan 2019
  
  %\cite{Ackermann:2015lka}
\bibitem{Ackermann:2015lka}
  M.~Ackermann {\it et al.} [Fermi-LAT Collaboration],
  ``Updated search for spectral lines from Galactic dark matter interactions with pass 8 data from the Fermi Large Area Telescope,''
  Phys.\ Rev.\ D {\bf 91} (2015) no.12,  122002,
 % doi:10.1103/PhysRevD.91.122002
  [arXiv:1506.00013 [astro-ph.HE]].
  %%CITATION = doi:10.1103/PhysRevD.91.122002;%%
  %195 citations counted in INSPIRE as of 21 Jan 2019
  
  
  %\cite{Gustafsson:2007pc}
\bibitem{Gustafsson:2007pc}
  M.~Gustafsson, E.~Lundstrom, L.~Bergstrom and J.~Edsjo,
  ``Significant Gamma Lines from Inert Higgs Dark Matter,''
  Phys.\ Rev.\ Lett.\  {\bf 99} (2007) 041301
  doi:10.1103/PhysRevLett.99.041301
  [astro-ph/0703512 [ASTRO-PH]].
  %%CITATION = doi:10.1103/PhysRevLett.99.041301;%%
  %233 citations counted in INSPIRE as of 10 Jun 2019
  
 %\cite{Garcia-Cely:2015khw}
\bibitem{Garcia-Cely:2015khw}
  C.~Garcia-Cely, M.~Gustafsson and A.~Ibarra,
  ``Probing the Inert Doublet Dark Matter Model with Cherenkov Telescopes,''
  JCAP {\bf 1602} (2016) no.02,  043
  doi:10.1088/1475-7516/2016/02/043
  [arXiv:1512.02801 [hep-ph]].
  %%CITATION = doi:10.1088/1475-7516/2016/02/043;%%
  %38 citations counted in INSPIRE as of 10 Jun 2019 
  
%\cite{Duerr:2015bea}
\bibitem{Duerr:2015bea}
  M.~Duerr, P.~Fileviez Pérez and J.~Smirnov,
  ``Gamma-Ray Excess and the Minimal Dark Matter Model,''
  JHEP {\bf 1606} (2016) 008
  doi:10.1007/JHEP06(2016)008
  [arXiv:1510.07562 [hep-ph]].
  %%CITATION = doi:10.1007/JHEP06(2016)008;%%
  %23 citations counted in INSPIRE as of 10 Jun 2019  
  
 %\cite{Duerr:2015aka}
\bibitem{Duerr:2015aka}
  M.~Duerr, P.~Fileviez Pérez and J.~Smirnov,
  ``Scalar Dark Matter: Direct vs. Indirect Detection,''
  JHEP {\bf 1606} (2016) 152
  doi:10.1007/JHEP06(2016)152
  [arXiv:1509.04282 [hep-ph]].
  %%CITATION = doi:10.1007/JHEP06(2016)152;%%
  %49 citations counted in INSPIRE as of 10 Jun 2019
  
%\cite{Jackson:2013pjq}
\bibitem{Jackson:2013pjq}
  C.~B.~Jackson, G.~Servant, G.~Shaughnessy, T.~M.~P.~Tait and M.~Taoso,
  ``Gamma-ray lines and One-Loop Continuum from s-channel Dark Matter Annihilations,''
  JCAP {\bf 1307} (2013) 021
  doi:10.1088/1475-7516/2013/07/021
  [arXiv:1302.1802 [hep-ph]].
  %%CITATION = doi:10.1088/1475-7516/2013/07/021;%%
  %34 citations counted in INSPIRE as of 10 Jun 2019  
  
  
%\cite{Jackson:2009kg}
\bibitem{Jackson:2009kg}
  C.~B.~Jackson, G.~Servant, G.~Shaughnessy, T.~M.~P.~Tait and M.~Taoso,
  ``Higgs in Space!,''
  JCAP {\bf 1004} (2010) 004
  doi:10.1088/1475-7516/2010/04/004
  [arXiv:0912.0004 [hep-ph]].
  %%CITATION = doi:10.1088/1475-7516/2010/04/004;%%
  %100 citations counted in INSPIRE as of 10 Jun 2019  
  
  
 %\cite{Bergstrom:1997fh}
\bibitem{Bergstrom:1997fh}
  L.~Bergstrom and P.~Ullio,
  ``Full one loop calculation of neutralino annihilation into two photons,''
  Nucl.\ Phys.\ B {\bf 504} (1997) 27
  doi:10.1016/S0550-3213(97)00530-0
  [hep-ph/9706232].
  %%CITATION = doi:10.1016/S0550-3213(97)00530-0;%%
  %262 citations counted in INSPIRE as of 10 Jun 2019 
   
%\cite{Bergstrom:2005ss}
\bibitem{Bergstrom:2005ss}
  L.~Bergstrom, T.~Bringmann, M.~Eriksson and M.~Gustafsson,
  ``Gamma rays from heavy neutralino dark matter,''
  Phys.\ Rev.\ Lett.\  {\bf 95} (2005) 241301
  doi:10.1103/PhysRevLett.95.241301
  [hep-ph/0507229].
  %%CITATION = doi:10.1103/PhysRevLett.95.241301;%%
  %116 citations counted in INSPIRE as of 10 Jun 2019  
  
  
%\cite{Cirelli:2015bda}
\bibitem{Cirelli:2015bda}
  M.~Cirelli, T.~Hambye, P.~Panci, F.~Sala and M.~Taoso,
  ``Gamma ray tests of Minimal Dark Matter,''
  JCAP {\bf 1510} (2015) no.10,  026
  doi:10.1088/1475-7516/2015/10/026
  [arXiv:1507.05519 [hep-ph]].
  %%CITATION = doi:10.1088/1475-7516/2015/10/026;%%
  %74 citations counted in INSPIRE as of 10 Jun 2019  
  
  
%\cite{Garcia-Cely:2015dda}
\bibitem{Garcia-Cely:2015dda}
  C.~Garcia-Cely, A.~Ibarra, A.~S.~Lamperstorfer and M.~H.~G.~Tytgat,
  ``Gamma-rays from Heavy Minimal Dark Matter,''
  JCAP {\bf 1510} (2015) no.10,  058
  doi:10.1088/1475-7516/2015/10/058
  [arXiv:1507.05536 [hep-ph]].
  %%CITATION = doi:10.1088/1475-7516/2015/10/058;%%
  %53 citations counted in INSPIRE as of 10 Jun 2019  
  
 %\cite{Duerr:2015vna}
\bibitem{Duerr:2015vna}
  M.~Duerr, P.~Fileviez Perez and J.~Smirnov,
  ``Gamma Lines from Majorana Dark Matter,''
  Phys.\ Rev.\ D {\bf 93} (2016) 023509,
%  doi:10.1103/PhysRevD.93.023509
  [arXiv:1508.01425 [hep-ph]].
  %%CITATION = doi:10.1103/PhysRevD.93.023509;%%
  %11 citations counted in INSPIRE as of 21 Jan 2019   
  
  
 %\cite{Bergstrom:2004nr}
\bibitem{Bergstrom:2004nr}
  L.~Bergstrom, T.~Bringmann, M.~Eriksson and M.~Gustafsson,
  ``Two photon annihilation of Kaluza-Klein dark matter,''
  JCAP {\bf 0504} (2005) 004
  doi:10.1088/1475-7516/2005/04/004
  [hep-ph/0412001].
  %%CITATION = doi:10.1088/1475-7516/2005/04/004;%%
  %86 citations counted in INSPIRE as of 10 Jun 2019 
  
%\cite{Bergstrom:2004cy}
\bibitem{Bergstrom:2004cy}
  L.~Bergstrom, T.~Bringmann, M.~Eriksson and M.~Gustafsson,
  ``Gamma rays from Kaluza-Klein dark matter,''
  Phys.\ Rev.\ Lett.\  {\bf 94} (2005) 131301
  doi:10.1103/PhysRevLett.94.131301
  [astro-ph/0410359].
  %%CITATION = doi:10.1103/PhysRevLett.94.131301;%%
  %224 citations counted in INSPIRE as of 10 Jun 2019
  
  
 %\cite{Ackermann:2013uma}
\bibitem{Ackermann:2013uma}
  M.~Ackermann {\it et al.} [Fermi-LAT Collaboration],
  ``Search for Gamma-ray Spectral Lines with the Fermi Large Area Telescope and Dark Matter Implications,''
  Phys.\ Rev.\ D {\bf 88} (2013) 082002,
%  doi:10.1103/PhysRevD.88.082002
  [arXiv:1305.5597 [astro-ph.HE]].
  %%CITATION = doi:10.1103/PhysRevD.88.082002;%%
  %253 citations counted in INSPIRE as of 21 Jan 2019 
  
 %\cite{Abramowski:2013ax}
\bibitem{Abramowski:2013ax}
  A.~Abramowski {\it et al.} [H.E.S.S. Collaboration],
  ``Search for Photon-Linelike Signatures from Dark Matter Annihilations with H.E.S.S.,''
  Phys.\ Rev.\ Lett.\  {\bf 110} (2013) 041301,
 % doi:10.1103/PhysRevLett.110.041301
  [arXiv:1301.1173 [astro-ph.HE]].
  %%CITATION = doi:10.1103/PhysRevLett.110.041301;%%
  %224 citations counted in INSPIRE as of 26 Jan 2019 
  
%\cite{Alioli:2017nzr}
\bibitem{Alioli:2017nzr}
  S.~Alioli, M.~Farina, D.~Pappadopulo and J.~T.~Ruderman,
  ``Catching a New Force by the Tail,''
  Phys.\ Rev.\ Lett.\  {\bf 120} (2018) no.10,  101801,
%  doi:10.1103/PhysRevLett.120.101801
  [arXiv:1712.02347 [hep-ph]].
  %%CITATION = doi:10.1103/PhysRevLett.120.101801;%%
  %12 citations counted in INSPIRE as of 23 Jan 2019  

%\cite{Acharya:2017ttl}
\bibitem{Acharya:2017ttl}
  B.~S.~Acharya {\it et al.} [Cherenkov Telescope Array Consortium],
  ``Science with the Cherenkov Telescope Array,''
  arXiv:1709.07997 [astro-ph.IM].
  %%CITATION = ARXIV:1709.07997;%%
  %75 citations counted in INSPIRE as of 24 Jan 2019  


%\cite{FileviezPerez:2018jmr}
\bibitem{FileviezPerez:2018jmr}
  P.~Fileviez Perez, E.~Golias, R.~H.~Li and C.~Murgui,
  ``Leptophobic Dark Matter and the Baryon Number Violation Scale,''
  arXiv:1810.06646 [hep-ph].
  %%CITATION = ARXIV:1810.06646;%%    
  
  
  
  %\cite{Duerr:2015wfa}
\bibitem{Duerr:2015wfa}
  M.~Duerr, P.~Fileviez Perez and J.~Smirnov,
  ``Simplified Dirac Dark Matter Models and Gamma-Ray Lines,''
  Phys.\ Rev.\ D {\bf 92} (2015) no.8,  083521,
%  doi:10.1103/PhysRevD.92.083521
  [arXiv:1506.05107 [hep-ph]].
  %%CITATION = doi:10.1103/PhysRevD.92.083521;%%
  %33 citations counted in INSPIRE as of 21 Jan 2019 
  
  
  
  
 %%%%%%%%%%%%%%%%%%%%%%%%%%%%%%%%%%%%%%%%%%%%%%%%%%%%%%%%%%%%%%%%%%%%%%%%%%%%%%%
\end{thebibliography}
\end{document}